\documentclass[prd, superscriptaddress, preprintnumbers, twocolumn, showpacs]{revtex4}
\usepackage{graphicx}
\usepackage{amsmath}
\usepackage{multirow}
\usepackage{bm}
\usepackage{color}
\usepackage{ulem}

\newcommand{\be}{\begin{equation}}
\newcommand{\ee}{\end{equation}}
\newcommand{\ba}{\begin{eqnarray}}
\newcommand{\ea}{\end{eqnarray}}

\newcommand{\tr}{\,\mbox{tr}}
\newcommand{\sign}{\,\mbox{sign}}

\begin{document}
\normalem

\title{Chiral asymmetry of the Fermi surface in dense relativistic matter in a magnetic field}
\date{\today}


\author{E. V. Gorbar}
\email{gorbar@bitp.kiev.ua}
\affiliation{Bogolyubov Institute for Theoretical Physics, 03680, Kiev, Ukraine}

\author{V. A. Miransky}
\email{vmiransk@uwo.ca}
\altaffiliation[On leave from ]{Bogolyubov Institute for Theoretical Physics, 03680, Kiev, Ukraine.}
\affiliation{Department of Applied Mathematics, University of Western Ontario, London, Ontario N6A 5B7, Canada}

\author{I. A. Shovkovy}
\email{igor.shovkovy@asu.edu}
\affiliation{Department of Applied Sciences and Mathematics, Arizona State University, Mesa, Arizona 85212, USA}

\begin{abstract}
It is revealed that in the normal phase of dense relativistic matter in a magnetic field, there 
exists a contribution to the axial current associated with a relative shift of the longitudinal 
momenta in the dispersion relations of opposite chirality fermions. Unlike the topological 
contribution in the axial current at the lowest Landau level, recently discussed in the 
literature, the dynamical one appears only in interacting matter and affects the fermions 
in {\em all} Landau levels, including those around the Fermi surface. The induced axial current 
and the shift of the Fermi surfaces of the left-handed and right-handed fermions are expected 
to play an important role in transport and emission properties of matter in various types of 
compact stars as well as in heavy ion collisions.
\end{abstract}


\maketitle

{\em Introduction.---}
At zero density and temperature, the structure of the ground state in relativistic chiral 
invariant theories in a magnetic field is dictated by the magnetic catalysis phenomenon: the magnetic field
is a strong catalyst of spontaneous chiral symmetry breaking \cite{magCat,magCat3+1}. The situation becomes much
more complicated in the case of dense relativistic matter in a magnetic field \cite{FI1}. The main goal of this
Rapid Communication is to reveal and describe some universal properties of such dynamics.

The recent studies \cite{GGM2007} of similar dynamics in graphene in $2 + 1$ dimensions
have revealed several types of order parameters whose analogs have not been discussed in 
the context of relativistic models in $3+1$ dimensions. (For earlier applications of the 
magnetic catalysis phenomenon in graphene, see Ref.~\cite{mc1}.) This motivated us to 
reexamine the properties of dense relativistic matter in an external magnetic field in 3+1 
dimensions. As we show, an external magnetic field induces two qualitatively different 
contributions to the net axial current in the normal phase of dense relativistic matter. 
The first one is a topological contribution {due to the lowest Landau level (LLL)} that 
was previously discussed in the literature \cite{Son:2004tq,Metlitski:2005pr,Charbonneau:2009ax,Rebhan}. 
It exists even in free theories. (For related studies in hot quark-gluon plasma, see 
Ref.~\cite{Kharzeev:2007tn}.) In this Rapid Communication we find that there is also an additional 
dynamical contribution to the axial current that appears only in interacting matter. 
The crucial new point is that {\em all} Landau levels, including those around 
the Fermi surface, contribute to this interaction-driven contribution. Its origin is 
related to a relative shift of the longitudinal momenta in the dispersion relations 
of opposite chirality fermions. The amount of the shift is proportional to a coupling 
constant, the magnetic field, and the fermion chemical potential. Notably, it is almost 
independent of temperature. As we discuss below, this effect may have profound implications 
for the physics of compact stars as well as heavy ion collisions.

{\em Model.---} 
{In this paper, in order to illustrate the effect in the clearest way, a simple 
model will be utilized. We perform a study of a Nambu-Jona-Lasinio model with 
the Lagrangian density}
\be
{\cal L}= \bar\psi \left(iD_\nu+\mu_0\delta_{\nu}^{0}\right)\gamma^\nu \psi
+\frac{G_{\rm int}}{2}\left[\left(\bar\psi \psi\right)^2
+\left(\bar\psi i\gamma^5\psi\right)^2\right],
\label{model}
\ee
where $\gamma^5\equiv i\gamma^0\gamma^1\gamma^2\gamma^3$ and $\mu_0$ is the chemical 
potential. This model possesses the chiral $U(1)_L\times U(1)_R$ symmetry. The covariant 
derivative $D_{\nu}=\partial_\nu -i e A_{\nu}$ is taken in the Landau gauge, i.e., 
$A_{\nu}= x B \delta_{\nu}^{2}$ where $B$ is the strength of the external 
magnetic field pointing in the $z$-direction.

The structure of the full fermion propagator is given by a (3+1)-dimensional 
generalization of the ansatz used for the description of the quantum Hall effect 
dynamics in graphene \cite{GGM2007}, namely
\ba
iG^{-1}(u,u^\prime) &=&\Big[(i\partial_t+\mu)\gamma^0 -
(\bm{\pi}\cdot\bm{\gamma})-\pi^3\gamma^3
\nonumber\\
&+& i\tilde{\mu}\gamma^1\gamma^2
+\Delta\gamma^3\gamma^5
-m\Big]\delta^{4}(u- u^\prime),
\label{ginverse}
\ea
where $\pi_k = i (\partial_k -i e A_{k})$ is the canonical momentum and $u=(t,\bm{r})$. 
The above propagator contains several dynamical parameters that are absent in the 
tree level propagator, 
\be
iS^{-1}(u,u^\prime) =\left[(i\partial_t+\mu_0)\gamma^0
-(\bm{\pi}\cdot\bm{\gamma})-\pi^3\gamma^3\right]\delta^{4}(u- u^\prime).
\ee
The physical meaning of the parameters $m$ and $\mu$ is straightforward: $m$ is 
the Dirac mass and $\mu$ is the full chemical potential in interacting theory. 
From the structure of the inverse propagator in Eq.~(\ref{ginverse}), it is clear 
that $\tilde{\mu}$ plays the role of the anomalous magnetic moment 
[in graphene, it also can be interpreted as a chemical potential 
related to a conserved (pseudospin) current \cite{GGM2007}].

The meaning of the last parameter, $\Delta$, is more subtle. In the context 
of graphene, $\Delta$ is a mass parameter that is odd both under time-reversal 
and parity transformations. In 2+1 dimensional models, this mass is responsible 
for inducing the Chern-Simons term in the effective action for gauge fields 
\cite{CS}. In 3+1 dimensions, as suggested by Eq.~(\ref{ginverse}), it is 
related to an induced axial current along the direction of the magnetic
field, $\bar\psi \gamma^3\gamma^5 \psi$. As will be shown below, $\Delta$
determines a shift of the longitudinal momenta in the dispersion relations of 
opposite chirality fermions in the chiral limit. We will call it a chiral shift 
parameter. Let us emphasize that, in the presence of an external magnetic field, 
the time-reversal and parity symmetries are broken and a state with the vanishing 
$\Delta$ is not protected by any symmetry.

The parameters $m$, $\mu$, $\Delta$ and $\tilde{\mu}$ are self-consistently 
determined from the gap equation, which takes the following form 
in the mean field approximation: 
\ba
G^{-1}(u,u^\prime) &=& S^{-1}(u,u^\prime)
- i G_{\rm int} \left\{ 
   G(u,u) 
-  \gamma^5 G(u,u) \gamma^5 \right.\nonumber\\
&-&\left. \mbox{tr}[G(u,u)] 
+  \gamma^5\, \mbox{tr}[\gamma^5G(u,u)]\right\}\delta^{4}(u- u^\prime).\nonumber\\
\label{gap}
\ea
While the parameters $\Delta$ and $\mu$ do not break the chiral 
$U(1)_L\times U(1)_R$ symmetry, nonzero values of $m$ and $\tilde{\mu}$ break 
it down to $U(1)_{L+R}$. In the mean field approximation used here, we find that 
$\tilde{\mu} = 0$ in a self-consistent solution to the gap equation. While this fact 
simplifies the analysis, we note that $\tilde{\mu}$ may be non-vanishing in more refined 
approximations and in models with other types of interactions \cite{GGM2007,magCatFI}. 
On the other hand, as suggested by the analysis in graphene, a nonzero $\tilde{\mu}$ 
should not change the main qualitative features of the phase with {an} induced 
$\Delta$ \cite{GGM2007}.

The spectrum of fermionic quasiparticles is determined by the poles in the full propagator 
(\ref{ginverse}) with $\tilde\mu=0$. As in Refs.~\cite{magCat3+1,GGM2007}, expanding 
the propagator over the Landau levels, we arrive at the following dispersion relation:
\be
\omega_{n,\sigma}=-\mu\pm\sqrt{\left[\sqrt{m^2+k_3^2}+ \sigma \Delta\sign(eB)\right]^2+2n|eB|},
\label{disp-relation}
\ee
where $n$ labels the Landau levels, $\sigma =\pm1$, and $k_3$ is the momentum along 
the direction of the magnetic field. In the chiral limit with $m=0$, the states with
the quantum number $\sigma =\pm1$ have a shifted longitudinal momentum, 
$k^3 \to k^3 \pm \Delta\sign(eB)$. As will be shown below, in this limit,
the two different values of $\sigma$ correspond to fermions with opposite chiralities. 
As a result, the Fermi surfaces for left-handed and right-handed fermions become shifted.

{\em Results.---}
In accordance with the magnetic catalysis scenario \cite{magCat,magCat3+1}, the ground 
state in the NJL model at vanishing $\mu_0$ is characterized by a nonzero Dirac mass 
$m_0$ that breaks the chiral $U(1)_L\times U(1)_R$ symmetry. Such a vacuum state can 
withstand a finite stress due to a nonzero chemical potential. However, when $\mu_0$ 
exceeds a certain critical value $\mu_c$, the chiral symmetry restoration and a new 
ground state are expected. As we show here, the new state is characterized by a 
non-vanishing chiral shift parameter $\Delta$ and a nonzero axial current in the 
direction of the magnetic field. Since no symmetry of the theory is broken, this state 
describes the {\em normal} phase of matter that happens to have a rather rich chiral 
structure.

The value of the Dirac mass $m_0$ in the vacuum state was calculated in Ref.~\cite{magCat3+1}. 
In the weakly coupled regime, $g\equiv G_{\rm int} \Lambda^2/(4\pi^2) \ll 1$, the solution 
reads
\be
m_0^2 =\frac{1}{\pi l^2}\exp\left(-\frac{\Lambda^2 l^2}{g}\right),
\label{DiracMass}
\ee
where $l=1/\sqrt{|eB|}$ is the magnetic length and $\Lambda$ is the ultraviolet 
cutoff in the model at hand. This zero temperature solution exists for $\mu_0< m_0$. 
In this solution, the full chemical potential $\mu=\mu_0$.

Our analysis shows that, in addition to the solution with a nonzero Dirac mass $m$, 
the gap equation also allows a solution with $m=0$ and a nonzero chiral shift parameter $\Delta$, 
\be
\Delta =g \mu\frac{eB}{\Lambda^2\left[1+2 a g\right]+g|eB|},
\label{Delta-vs-mu}
\ee
where $a$ is a dimensionless constant of order $1$, which is determined by the 
regularization scheme used in the analysis (see below). Interestingly, {
the temperature dependence of $\Delta$ comes only through the chemical potential, 
which has a weak temperature dependence when $T\ll\mu$ \cite{GMS2009}. 
At $T=0$, the chemical potential satisfies the following equation:}
\ba
\mu &=& \mu_0-\frac{g}{(\Lambda l)^2}\left[\mu-\Delta\sign (eB)\right]\nonumber\\
&-&\frac{2g\sign (\mu)}{(\Lambda l)^2} \sum_{n=1}^{\infty} 
\sqrt{\mu^2-2n|eB|}\,\theta\left(\mu^2-2n|eB|\right).
\ea
The approximate solution to this equation is $\mu\simeq\mu_0$ up to power corrections 
in small $g$. 

The free energies of the two states are given by \cite{GMS2009}
\be
\Omega_{m} \simeq  -\frac{m_0^2}{2(2\pi l)^2}
\left(1+(m_0 l)^2\ln|\Lambda l|\right)
\label{Omega0}
\ee
and 
\be
\Omega_{\Delta} \simeq -\frac{\mu_0^2}{(2\pi l)^2}\left(1-g\frac{|eB|}{\Lambda^2}\right),
\label{OmegaDelta0}
\ee 
respectively. In deriving the last expression, we used the approximate relations 
$\mu\simeq \mu_0$ and $\Delta \simeq g\mu_0 eB/\Lambda^2$. By comparing the free 
energies in Eqs.~(\ref{Omega0}) and (\ref{OmegaDelta0}), we see that the ground 
state with a nonzero $\Delta$ becomes favorable when $\mu_0\gtrsim m_0/\sqrt{2}$. 
This is analogous to the Clogston relation in superconductivity \cite{Clogston}. 

The properties of the two types of quasiparticles corresponding to $\sigma = \pm 1$
in the dispersion relation in Eq.~(\ref{disp-relation}) with $m = 0$ are further 
clarified by the structure of the full propagator:
\be
G(u,u)=G_0^{-}{\cal P}_{-}
+\sum_{n=1}^{\infty}\left(G_{n}^{-}{\cal P}_{-}+G_{n}^{+}{\cal P}_{+}\right),
\label{full-propagator}
\ee
where ${\cal P}_{\pm}\equiv \frac12 \left[1\pm i\gamma^1\gamma^2\sign (eB)\right]$ are 
the spin projectors. (For $u^\prime\neq u$, the propagator will be presented elsewhere 
\cite{GMS2009}.) The functions $G_{n}^{\pm}$ with $n\geq 0$ are given by
\ba
G_{n}^{\pm} &=& 
\frac{i|eB|\gamma^0}{2\pi}\int\frac{d\omega d k^3}{(2\pi)^2} \nonumber\\
&&\times \Big[
 \frac{\omega+\mu \pm[k^3-\Delta \sign(eB)]}
      {(\omega+\mu)^2 - 2n|eB| - [k^3- \Delta\sign(eB)]^2 }{\cal P}_{5}^{-}
\nonumber\\
&&+\frac{\omega+\mu \mp[k^3+\Delta \sign(eB)]}
      {(\omega+\mu)^2 - 2n|eB| - [k^3+ \Delta\sign(eB)]^2 }{\cal P}_{5}^{+}
\Big],\nonumber\\
\label{G_n}
\ea
where ${\cal P}_{5}^{\pm}\equiv \frac12 \left[1\pm \gamma^5\sign (eB)\right]$ are the 
chirality projectors for a fixed sign of $eB$. As follows from this equation, the 
quasiparticles of opposite chiralities have dispersion relations that differ from 
those in the free theory by the shift of their longitudinal momentum 
$k^3\to k^3\pm\Delta\sign(eB)$. This has profound implications for the physical 
properties of matter.

The ground state with $\Delta\neq 0$ is characterized by a non-vanishing expectation 
value of the axial current density, 
\ba
\langle j_5^3(u)\rangle &=& -\tr\left[\gamma^3\gamma^5 G(u,u)\right]
=\frac{eB}{2\pi^2}\mu-\frac{|eB|}{2\pi^2}\Delta \nonumber\\
&-&\frac{|eB|}{\pi^2}\Delta \sum_{n=1}^{\infty} \kappa(\sqrt{2n|eB|},\Lambda) ,
\label{j5}
\ea
where $\kappa(x,\Lambda)$ is a smooth cutoff function defined by the value of the 
cut-off energy $\Lambda$ and a certain width of the region in which the value of the 
function drops from $1$ down to $0$ (i.e., $\kappa(x,\Lambda)\simeq 1$ for $x\ll \Lambda$
and $\kappa(x,\Lambda)\simeq 0$ for $x\gg \Lambda$). Taking this into account, we 
find that $\sum_{n=1}^{\infty} \kappa(\sqrt{2n|eB|},\Lambda) = a \Lambda^2/|eB|$ 
and $a=O(1)$.

While the first two terms in the current density (\ref{j5}) come from {the LLL}, 
the last term is due to the higher Landau levels, $n\geq 1$. Notably, all higher 
Landau levels below the (smeared) cut-off energy give (nearly) identical contributions 
to the induced axial current. The contribution proportional to $\mu$ is topological in 
nature and appears even in the free theory \cite{Metlitski:2005pr}. All other terms, 
which are proportional to $\Delta$, are the result of interactions and have not been 
{revealed} in the literature before. 

We expect that the interaction-driven contributions can strongly modify transport 
and emission properties of dense relativistic matter in a magnetic field. Indeed, the 
corresponding induced axial currents are the consequence of a relative longitudinal flow 
of opposite chirality quasiparticles, including those in higher Landau levels around the 
Fermi energy. This is in contrast to the role of the topological contribution that is 
exclusively due to {the LLL}, which is to a large degree quenched from the low-energy 
dynamics by the Pauli exclusion principle in many realistic cases. 

{\em Discussion and Summary.---}
In this Rapid Communication we found that, in accordance with the magnetic catalysis scenario, the vacuum 
state of relativistic matter in a magnetic field is characterized by a non-zero Dirac mass 
given by Eq.~(\ref{DiracMass}) at weak coupling. However, when the chemical potential 
exceeds a certain critical value, $\mu_c\simeq m_0/\sqrt{2}$, such a state is replaced 
by the normal ground state that is characterized by the following two properties: 
(i) the presence of an induced axial current along the magnetic field and 
(ii) the presence of the dynamically generated chiral shift parameter $\Delta$, which 
is a $3 + 1$ dimensional analog of the parity odd mass term in $2 + 1$ dimensions leading 
to the Chern-Simons term. We find that, in addition to the previously known topological term 
in the induced current, there are also interaction-driven contributions from the lowest as 
well as from the higher Landau levels. In fact, the newly found contributions are directly 
related to a dynamically generated value of the chiral shift parameter 
$\Delta\simeq g\mu_0 eB/\Lambda^2$. This parameter quantifies the relative shift of the 
longitudinal momenta in the dispersion relations of opposite chirality quasiparticles.

{
It might be appropriate to mention that, for instructional purposes, our study here 
is simplified: we used an NJL-type local interaction and utilized the mean-field 
approximation. These limitations may lead to the results that are not always quantitatively 
reliable, e.g., in the context of stellar matter. Nevertheless, it is expected that our 
results should remain qualitatively the same even when more realistic models are used.
Indeed, the fact that the expression for $\Delta$ in Eq.~(\ref{Delta-vs-mu}) is linear 
in $g$ in the lowest order indicates that the 
corresponding dynamics is essentially perturbative. Apparently this is a general feature 
that should not depend on whether the interactions are short range, as in the NJL model, 
or long range, as in QCD or QED. In either case, a vanishing $\Delta$ is not protected by 
any symmetry.}

Another limiting assumption of this study is the exact chiral symmetry. However, most of 
the results are not modified much when the symmetry is at least approximate, i.e., 
when the fermions have non-zero bare Dirac masses, but such masses are small compared 
to the value of the chemical potential \cite{GMS2009}. {For the applications 
in protoneutron stars suggested below, this approximation is justified, but the general
study will be of interest, e.g., in relation to the electron plasma in white dwarfs.}

In the future, it will be also of interest to address a possible {interference} 
of the dynamics responsible for the generation of the chiral shift parameter with color 
superconductivity in quark matter \cite{Alford:2007xm}. Here we just mention that the 
normal ground state with a nonzero $\Delta$ seems to be the only possibility at 
{higher temperatures, which are of main interest for us here.}

We expect that the generation of the chiral shift parameter may affect physical 
properties of the quark matter in quark and/or hybrid stars, the electron gas in 
neutron stars, and possibly even the electron gas in white dwarfs. The corresponding
fermionic systems are degenerate ($T\ll\mu$) and the results of this Rapid Communication apply 
directly. It is appropriate to mention, however, that the chiral shift parameter does 
not vanish even in the non-degenerate limit ($T\gg\mu$), although the analysis of the 
dynamics becomes more involved \cite{GMS2009}. Therefore, the generation of a 
non-zero $\Delta$ can also affect the chiral magnetic effect in heavy ion collisions 
\cite{Kharzeev:2007tn}.

One of the consequences of the phenomenon discussed in this Rapid Communication is the possibility 
of a qualitatively new mechanism for the pulsar kicks \cite{pulsarkicks}. In the presence 
of a magnetic field, almost any type of relativistic matter inside a protoneutron star 
should develop axial currents as in Eq.~(\ref{j5}). The main carriers of such currents
are the electrons in the nuclear matter, and the quarks together with the electrons in 
the deconfined quark matter. Since the induced currents and the chiral shift parameter 
have only a weak temperature dependence (assuming $T\ll\mu$), {this phenomenon} may 
provide a robust anisotropic medium even at the earliest stages of the protoneutron 
star. This is important because {moderately} hot matter with {$10~\mbox{MeV}
\lesssim T\lesssim 50~\mbox{MeV}$, present during the first few tens of seconds of the 
protoneutron star evolution \cite{AnnRevNuclPart},} may have a large enough amount 
of the thermal energy to power the strongest {(with $v\gtrsim 1000~\mbox{km/s}$)} 
pulsar kicks observed \cite{pulsarkicks}. In contrast, the constraints of the energy 
conservation make it hard, if not impossible, to explain such kicks if the interior 
matter is cold ($T\lesssim 1~\mbox{MeV}$). The common difficulty of using a hot matter, 
however, is the very efficient thermal isotropization that washes out a non-isotropic 
distribution of neutrinos produced by almost any mechanism \cite{Kusenko,SagertSchaffner}. 
In the mechanism proposed in this Rapid Communication, however, the asymmetric distribution of the 
neutrinos {\em develops} as a result of their multiple elastic scattering on the left-handed 
electrons or quarks, flowing in the whole bulk of the stellar matter in one direction 
along the magnetic field. 

In passing, let us mention that the robustness of the axial currents in hot magnetized 
matter may also provide an additional neutrino push to facilitate successful supernova 
explosions as suggested in Ref.~\cite{Fryer:2005sz}. Further details of this mechanism 
will be discussed elsewhere \cite{GMS2009}. 

{\em Acknowledgments.---}
The authors would like to thank Valery Gusynin for fruitful discussions. {They also 
thank J.~Noronha, A.~Schmitt and D.~T.~Son for useful comments on the manuscript.} 
V.A.M. acknowledges fruitful discussions with Dima Kharzeev and Eric Zhitnitsky 
concerning the results in Refs.~\cite{Metlitski:2005pr,Kharzeev:2007tn}.
The work of E.V.G. was supported by the State Foundation for Fundamental
Research under the grant F/16-457-2007, and by the Program of Fundamental 
Research of the Physics and Astronomy Division of the National Academy of 
Sciences of Ukraine. The work of V.A.M. was supported by the Natural 
Sciences and Engineering Research Council of Canada. The work of I.A.S. 
was supported in part by the start-up funds from the Arizona State University.

\end{document}